# Stochastic Modeling of Filtration with Sieving in Graded Pore Networks

**B. Gu[1], P. Sanaei[2], L. Kondic[3], L.J. Cummings[3]**

[1]Department of Mathematical Sciences, Worcester Polytechnic Institute, Worcester, MA 01609, USA

[2]Department of Mathematics, Georgia State University, Atlanta, GA 30302, USA

[3]Department of Mathematical Sciences, New Jersey Institute of Technology, Newark, NJ 07102, USA
**Corresponding author:** Binan Gu, bgu@wpi.edu

We model filtration of a feed solution, containing both small and large foulant particles, by a membrane filter. The membrane interior is modeled as a network of pores, allowing for the simultaneous adsorption of small particles and sieving of large particles, two fouling mechanisms typically observed during the early stages of commercial filtration applications. In our model, first-principles continuum partial differential equations model transport of the small particles and adsorptive fouling in each pore, while sieving particles are assumed to follow a discrete Poisson arrival process with a biased random walk through the pore network. Our goals are to understand the relative influences of each fouling mode and highlight the effect of their coupling on the performance of filters with a pore-size gradient (specifically, we consider a banded filter with different pore sizes in each band). Our results suggest that, due to the discrete nature of pore blockage, sieving alters qualitatively the rate of the flux decline. Moreover, the difference between sieving particle sizes and the initial pore size (radius) in each band plays a crucial role in indicating the onset and disappearance of sieving-adsorption competition. Lastly, we demonstrate a phase transition in the filter lifetime as the arrival frequency of sieving particles increases.

## 1. Introduction

Membrane filtration of fluids is ubiquitous in industrial, biological, and environmental applications. The growing complexity of membrane designs and operating conditions has driven increased interest in mathematical models that can accurately capture the dynamics of flow and fouling within these systems. Many modeling approaches draw from the broader literature on porous media, where the internal geometry of a material is represented in simplified, often idealized forms to study transport phenomena. A particularly successful class of models treats the pore space as a network, an idea that dates back to the seminal work of Fatt (1956*a*,*b*,*c*). Since then, network-based models have become widespread across domains including fractured rocks (Sahimi 1993), oil reservoirs (Øren *et al.* 1998; Mehmani *et al.* 2013; Cui *et al.* 2022), $CO_2$ sequestration (Iglauer *et al.* 2015), hydrogen storage (Hashemi *et al.* 2021; Lysyy *et al.* 2024; Zhao *et al.* 2024), microvascular transport (Chang & Roper 2019) and leaf networks (Katifori *et al.* 2010; Katifori 2018). These models balance geometric fidelity with computational efficiency, making them especially attractive for simulating flow and particle transport in complex porous systems.

In parallel, imaging and data-processing techniques for characterizing the topology and connectivity of porous structures have advanced significantly. These include workflows such as micro-CT image acquisition and pre-processing (Blunt *et al.* 2013), voxel- or pixel-based network extraction methods (such as the watershed segmentation method used by Gerke *et al.* (2020) and a genetic-algorithm based approach in Ebrahimi *et al.* (2013)), and experimental measurement techniques (Ho & Zydney 2000). These developments have greatly facilitated the numerical investigation of fluid flow and transport phenomena within pore-scale network models (see, e.g., Blunt *et al.* (2002) in the context of multiphase flows).

Among these industrial contexts, membrane filtration has emerged as a field of growing interest in terms of modeling and simulations. Research to date has focused on two primary aspects, the first being appropriate representation of the spatial structure of membrane filters. This has been successfully modeled as multilayered porous media, represented either as void spaces between material matrices (Griffiths *et al.* 2016; Shirataki & Wickramasinghe 2023) or as interconnected pore networks (Griffiths *et al.* 2014; Sanaei & Cummings 2018; Gu *et al.* 2020; Chen *et al.* 2021; Fong *et al.* 2021; Gu *et al.* 2022). Extensions of these models introduce porosity or pore radius gradients, designed to maximize filter usage and prolong operational lifetime (Dalwadi *et al.* 2015; Griffiths *et al.* 2020; Gu *et al.* 2023). Dalwadi *et al.* (2015) consider fluid and particle dynamics in a 2D domain with arrays of impermeable obstacles (representing filter material) decreasing in size in the direction of flow, effectively generating a porosity gradient using the void space between the obstacles. Meanwhile, Griffiths *et al.* (2020), employing a network of pore throats and junctions, introduce a porosity gradient by skewing the initial distribution of the pore junctions: the denser the pore junctions, the higher the porosity. Gu *et al.* (2023) consider a similar network, but with a pore radius gradient instead of a porosity gradient, achieved by dividing the computational domain into several horizontal bands perpendicular to the direction of flow; specifically, porosity is fixed across all bands, and a cascade of pore radii is assigned to consecutive bands. Such design considerations influence the membrane's internal geometry and, in turn, affect topological descriptors of the pore network such as connectivity, coordination number, and tortuosity (Griffiths *et al.* 2020; Gu *et al.* 2022).

Secondly, appropriate modeling of membrane fouling has also received significant attention. The material properties of the porous material and the operating conditions of the filter give rise to multiple fouling mechanisms that result from foulant-material interactions: (1) adsorption (or standard blocking), a slow accretion of particles much smaller than pore sizes onto the pore walls; (2) sieving, a more abrupt blockage of pore pathways by particles comparable in size to the pores; and (3) caking, the buildup of a fouling layer on the membrane surface in later stages. These mechanisms directly impact filter performance and efficiency, typically quantified via measures such as the throughput of filtrate through the filter, the quantity of foulant retained by the filter, and the overall filter lifetime.

Adsorption is frequently the sole fouling mechanism modeled (considered by many of the aforementioned authors), in view of its early onset in most filtration processes and the strong effect it has on filter efficiency and function. Sieving has been less considered from a modeling perspective, but can also be important. Applications of sieving, in addition to membrane filtration, include biological processes such as nutrient sieving in bone joints (Steck & Tate 2005), renal filtration (Lawrence *et al.* 2017), food processing techniques such as milling and differential sieving (Sultanbawa *et al.* 2001; Cheng *et al.* 2022) and chemical treatments such as the carbon molecular sieve (Liu *et al.* 2023*b*).

An accurate theoretical description of sieving requires additional assumptions about how particles arrive at the filter upstream surface and how they then navigate the pore geometry. These assumptions vary across deterministic and stochastic modeling approaches. Kelly *et al.* (2023) model an experimental setup in which a suspension of colloidal particles passes under pressure through a porous medium comprising randomly distributed glass beads that form a stochastic network of channels and employ a deterministic suspension arrival process with constant rate. Their study focuses particularly on how driving pressure changes particle deposition patterns through the depth of the porous medium. Sanaei & Cummings (2017) provide one of the first few mathematical models that capture the effect of adsorption and sieving concurrently on filtration dynamics in “track-etched” type filters, where each pore forms an isolated path across the membrane (thus not a network model). There, adsorption is modeled as a sink for foulant concentration as it traverses the pore, which shrinks due to mass conservation, while sieving is represented as a partial differential equation that governs the probability of blocking events per unit area per time and is incorporated in the total resistance of the pore. Beuscher (2010) considers sieving in parallel layers of cylindrical pores and models particle transport as a $k$-nearest-neighbor random walk between pores in adjacent layers where $k$ is a function of lateral flow between pores. We refer interested readers to the book by Sahimi (1994) for further relevant theoretical discussions, such as the classical theory of percolation and the exclusion process (to which sieving is analogous), and to Majumdar *et al.* (1998) for adsorption and desorption on lattice models.

Building on these modeling efforts, we present a network-based model for simulating the fouling of pore-radius-graded membranes under combined sieving and adsorption. In our model, sieving is represented as a two-stage stochastic process on the pore network: first, particle arrival at the membrane surface is governed by a Poisson process; second, transport through the membrane follows a fluid flux-weighted random walk over network vertices, with sieving events recorded when particles reach pores too small to enter. Adsorption is modeled as a continuous decay in local permeability along each pore throat due to foulant deposition (see, for example, Sanaei & Cummings (2017)), allowing the model to capture both gradual and sudden reductions in fluid flux and throughput. The model supports networks of arbitrary topology and pore-radius gradients, enabling us to investigate the effects of membrane geometry and sieving particle size on key measures of performance, such as total throughput, foulant concentration in the filtrate and filter lifetime.

This paper is organized as follows. In section 2.1, we describe the generation of the pore-radius-graded network. We present the fluid flow model at the pore scale in section 2.2, and address adsorptive transport and pore radius evolution in section 2.3. In section 2.4, we introduce the sieving mechanism and underlying assumptions. We then nondimensionalize the model and derive the governing network dynamics in section 4. In section 5, we outline the numerical methods and investigation strategies, and then present simulation results that highlight the influence of key parameters on filter performance. We conclude in section 6 with a summary of findings and directions for future work.

## 2. Mathematical Modeling

In this section, we introduce the mathematical model in three stages. First, we outline the random network generation protocol following Gu *et al.* (2023), emphasizing its geometric constraints and their relevance to practical applications. We define the radius gradient and discuss how these constraints influence the density of pore junctions and throats within a given domain. Second, we formulate a first-principles model describing fluid flow, adsorptive foulant transport and pore size kinematics, beginning with a single pore and extending to a network of interconnected pores. Third, we develop the modeling assumptions for sieving–the central focus of this work–and examine their impact on fluid dynamics.

### 2.1. *Membrane Pore Network Generation*

The present work focuses on understanding the effect of sieving (pore blocking) within the interior of the pore network; hence, we will generate networks where the pore size decreases in the depth of the filter. We follow the approach of Gu *et al.* (2023), but condense and improve the presentation of that work below. We consider a metric graph $G = (V, E, D)$ where $V$ is the set of vertices and $E$ the set of edges (which will be expanded as circular cylinders to form pores) formed by a connection metric $D$. To generate the appropriate vertex set $V$, we first divide a rectangular prism (in $\mathbb{R}^3$) of square cross-section with side length $W$ and a height of $2W$ into $(m+2)$ horizontal bands, with uppermost and lowermost bands of thickness $W/2$, and $m$ additional interior bands of equal thickness $W/m$ (see Fig. 1). In previous work, porosity, defined as the total pore volume fraction, was found to be a property that strongly influences membrane performance (Gu *et al.* 2022); to control for its effect here, we will hold porosity fixed across all bands (a vertex density gradient across the bands is then inevitable).

We use $m = 4$ to illustrate our procedure (easily generalizable to arbitrary $m$). First, place uniformly randomly $N_k$ nodes in the interior bands for $k = 1, 2, 3, 4$, increasing from the inlet side to the outlet side of the membrane. We also place $N_{\text{in}} = 2N_1$ and $N_{\text{out}} = 2N_4$ nodes in the upstream and downstream outer bands, respectively (see Gu *et al.* (2023) for details on how $N_k$ is estimated for each band). The factor of 2 for the outermost bands is to ensure a suitable number of pores when the rectangular prism is cut to generate inlets and outlets (a procedure detailed below). Altogether, these nodes form the initial vertex set, $V_0$.

We define our connection metric $D$ as follows: connect the points $x_i, x_j \in V_0$ that lie in the spherical shell

$$D_{\min} < \left\| x_i - x_j \right\|_2 < D_{\max}\,, \tag{2.1}$$

where $\|\cdot\|_2$ is the Euclidean 2-norm. We also allow points to be connected through the four lateral sides of the prism, which is equivalent to enforcing a periodic boundary condition. Parameters $D_{\max}$ and $D_{\min}$ control the maximum and minimum pore length (discussion on how variations influence computational results may be found in Gu *et al.* (2022) and references therein). We generate membrane inlets and outlets by cutting the rectangular prism with two planes at a height of $0.5W$ and $1.5W$, respectively, producing a cube of side length $W$; the intersections of the cutting planes and the edges they cut naturally form the set of inlets $V_{\text{in}}$ on the membrane top surface, and outlets $V_{\text{out}}$ on the bottom surface. The vertices below $0.5W$ and above $1.5W$ are discarded from $V_0$, and the new vertices (inlets and outlets) generated by the cutting process are added, creating the final vertex set, $V$. The remaining vertices in the cube interior form the set of interior vertices, $V_{\text{int}} = V \setminus (V_{\text{in}} \cup V_{\text{out}})$. The edges/pores connected to the membrane's top and bottom surfaces are called boundary edges/pores. Because of this cutting procedure, each boundary edge connects a unique pair of an inlet/outlet and an interior vertex. We note in passing that under this protocol, the average number of adjacent edges for an interior vertex (coordination number in the industrial literature) has a range of $10-13$, which is representative of membrane filter materials at our considered pore scale (Hormann *et al.* 2016; Liu *et al.* 2023*a*; Zhu *et al.* 2024).

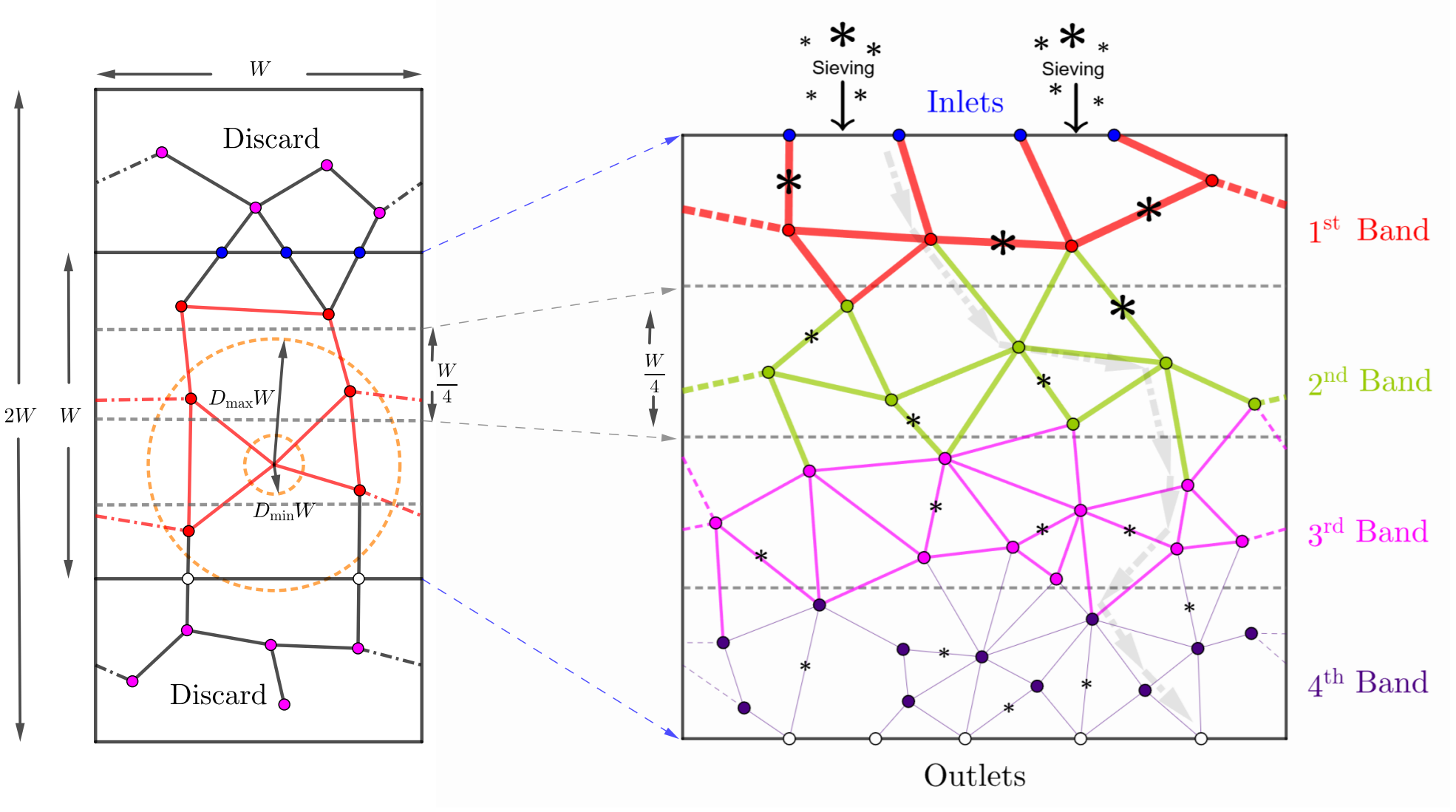

Figure 1. (Left) A visualization of the network generation protocol. (Right) Schematic of a 3D banded network represented in 2D. Colored junctions and pores correspond to each band as follows: red 1st band; green 2nd band; magenta 3rd band and indigo 4th band. Blue dots are inlets. White dots are outlets. Dashed lines are pores created by the periodic boundary conditions. The path connected with lightly shaded arrows is an example of a (possibly altered if without sieving) flow path when the network is sieved at some part in all bands by particles of different sizes (labeled by the asterisks).

Each edge $e_{ij} = (v_i, v_j)$ is associated with several relevant weights, such as pore radius $R_{ij}$ and pore length $L_{ij}$. To construct networks with a pore-radius gradient (see figure 1 for visualization of such a network that incorporates the details described here), we initially assign the same radius to each pore in the $k$-th band

$$R_n = R_m + (m-n)sW, \quad 1 \leq n \leq m, \tag{2.2}$$

where $s$ is the radius gradient/slope through the bands. For pores crossing two bands, we assign their radii according to the band in which they have more than half their lengths (see, for example, in the right panel of figure 1, the pore in red in the top left corner crossing from the 1st to the 2nd band is assigned to band 1). This rule of band assignment for each edge generates a band-specific set of edges $E_n$ such that $E = \bigcup_{n=1}^{m} E_n$.

In order to compare membrane networks on common grounds, we consider the following constraints on the band radii. First, we prescribe an average radius (taken over the bands) that provides a relationship between the radius gradient $s$ and the radius in the $m$-th (bottom) band

$$(\textit{Constraint 1}) \qquad R_0 := \frac{1}{m}\sum_{n=1}^{m} R_n = R_m + \frac{sW}{2}(m-1)\,. \tag{2.3}$$

Next, we insist that the porosity of each band is within a small relative tolerance $\delta$ (the value $\delta = 0.005$ is used in this work) of a prescribed porosity $\Phi_0$. This imposes a constraint on the number of vertices initially placed in each band, implemented via

$$(\textit{Constraint 2}) \qquad \frac{|\Phi_n - \Phi_0|}{\Phi_0} < \delta, \tag{2.4}$$

where $\Phi_n = \frac{\frac{\pi}{2}\sum_{e_{ij}\in E_n} R_{ij}^2 L_{n,ij}}{W^3/m}$ is the porosity of the $n$-th band, and $L_{n,ij}$ the length of the edge $e_{ij}$ inside the $n$-th band. We choose $\Phi_0 = 0.6$, corresponding to the porosity of typical commercial filter materials. The porosity-fixing procedures can be carried out in a predict-correct fashion, by either adding nodes when porosity overshoots or deleting nodes (and their connections) when porosity undershoots.

A consequence of fixing band porosities in a pore-size-graded network is that the larger the radius gradient, the larger the gradient of vertex/pore junction density. Furthermore, fulfilling *Constraint 2* also fixes the overall porosity of the entire membrane network via the relationship

$$\Phi = \frac{1}{m}\sum_{n=1}^{m} \Phi_n \approx \Phi_0. \tag{2.5}$$

We refer the reader to Gu *et al.* (2023) for more details on the generation of pore-size-graded networks and porosity-fixing techniques.

## 2.2. *Flow in an Edge and a Network of Pores*

In this section, we describe how a Newtonian fluid of viscosity $\mu$ flows through a single edge (pore). We use the Hagen-Poiseuille equation to relate fluid flux $Q_{ij}$ through each edge $(v_i, v_j) \in E$ with pore radius $R_{ij}$, pore length $L_{ij}$, and the pressure difference $P(v_i) - P(v_j)$ across adjacent vertices $v_i, v_j \in V$. More precisely,

$$Q_{ij} = K_{ij}\left(P(v_i) - P(v_j)\right), \quad (v_i, v_j) \in E, \tag{2.6}$$

where $K_{ij}$ are the entries of a diagonal matrix $K$ listing the *conductance* of each edge,

$$K_{ij} = \begin{cases} \frac{\pi R_{ij}^4}{8\mu L_{ij}}, & (v_i, v_j) \in E, \\ 0, & \text{otherwise.} \end{cases} \tag{2.7}$$

To ensure the Hagen-Poiseuille approximation remains valid we require that the aspect ratio of each edge $R_{ij}/L_{ij}$ is small, enforced by choosing $D_{\min}$ (minimum pore length in equation (2.1)) sufficiently large that

$$\frac{R_{ij}}{L_{ij}} \le \frac{\max\limits_{(v_i,v_j)\in E} R_{ij}}{D_{\min}} \ll 1.$$

Solving equation (2.6) on a graph with many edges simultaneously amounts to imposing conservation of fluid flux at each interior vertex. Doing so yields a system of algebraic equations for the pressures at interior vertices, known as the graph Laplace equation,

$$L_K P(v) = 0, \quad P(v) = \begin{cases} P_0, & v \in V_{\text{in}}, \\ 0, & v \in V_{\text{out}}, \end{cases} \tag{2.8}$$

where the graph Laplacian is defined by

DEFINITION 2.1. *(Graph Laplacian) The W-weighted graph Laplacian is given by*

$$L_{\mathcal{W}} := \mathcal{D} - \mathcal{W}, \tag{2.9}$$

*where $\mathcal{D}$ is the (diagonal) $\mathcal{W}$-weighted degree matrix for G, with entries*

$$\mathcal{D}_{ij} = \begin{cases} \sum_{k=1}^{|V|} \mathcal{W}_{ik}, & j = i, \\ 0, & \textit{otherwise}, \end{cases} \tag{2.10}$$

*and $\mathcal{W}$ is a weighted adjacency matrix, with nonnegative entries $\mathcal{W}_{ij}$ when $(v_i, v_j) \in E$, to be specified according to context. We also say that $e_{ij}$ is an edge connecting $v_i$ and $v_j$ iff $\mathcal{W}_{ij} > 0$ (or simply, $v_i$ and $v_j$ are connected or adjacent).*

We refer the reader to Gu *et al.* (2022) for a thorough description of the graphical setup and to Grady (2010) for a more comprehensive description of the function spaces endowed on graphs and technical details of calculus on graphs.

### 2.3. *Foulant Advection and Adsorptive Fouling*

The previous section was concerned solely with the flow of Newtonian fluid through the network. In this section, we address the fact that in filtration, the fluid is a feed solution carrying particles, which are removed by the filter, leading to fouling. Fouling can occur via a number of distinct mechanisms of which we consider two: adsorptive fouling by particles much smaller than pores, which are transported through the network by the flow and deposit on pore walls; and sieving by particles of size comparable to a typical pore radius. We first provide a model for adsorptive particle transport and fouling, and further discuss sieving in the next section.

We employ a continuum model for the foulant concentration within the feed, similar to those employed by Sanaei & Cummings (2018) and Gu *et al.* (2020). For each edge $e_{ij} = (v_i, v_j)$ of length $L_{ij}$, with $Y$ a local coordinate measuring distance along the edge from $v_i$, let $C_{ij}(Y, T)$ be the foulant particle concentration at any point of the edge at time $T$, then

$$Q_{ij}\frac{\partial C_{ij}}{\partial Y} = -\Lambda R_{ij}C_{ij}, \quad 0 \le Y \le L_{ij}, \tag{2.11a}$$

$$C_{ij}(0,T) = C_i(T), \quad (v_i, v_j) \in E, \tag{2.11b}$$

where $\Lambda$ is a parameter (with dimensions of velocity) that captures the affinity between the pore wall and foulant. Every pore junction $v_i \in V \backslash V_{\text{in}}$ is associated with a foulant concentration $C_i(T)$ that enforces flux balance, while the concentration at all inlets is the prescribed value $C_0$ in the incoming feed. The array of $C_i(T)$ values forms a vector $\boldsymbol{C}$.

First, we observe that equations (2.11$a$) and (2.11$b$) have an analytical solution

$$C_{ij}(Y,T) = C_i(T)\exp\left(-\frac{\Lambda\int_0^Y R_{ij}(Z,T)\,dZ}{Q_{ij}}\right), \quad 0 \le Y \le L_{ij}. \tag{2.12}$$

This formula helps express the foulant concentration at pore inlet $Y = 0$ and outlet $Y = L_{ij}$, respectively. It allows us to conveniently enforce conservation of particle flux at each vertex $v_i$, from which we derive a graph-transport equation for foulant concentration on the graph vertices, compactly written as (Gu *et al.* 2022)

$$L_Q^{\text{in}}\boldsymbol{C} = (Q \circ B)^{\text{T}}\boldsymbol{C}_0, \quad \forall T \ge 0, \tag{2.13}$$

$$\boldsymbol{C}_0(T) = (C_0, \ldots, C_0, 0, \ldots, 0)^{\text{T}}, \tag{2.14}$$

with boundary condition $C_0$ specified for each $v \in V_{\text{in}}$ (the number of nonzero entries $C_0$ in $\boldsymbol{C}_0$ is equal to $|V_{\text{in}}|$).

We briefly explain the terms in equation (2.13). The linear operator $L_Q^{\text{in}}$ is the in-degree weighted graph Laplacian, using $Q$ as a weight matrix to indicate directionality (see Gu *et al.* (2022) for implementation details and Chapman & Mesbahi (2011) for its conceptual origin from consensus dynamics), defined as

$$L_Q^{\text{in}} := D_{Q^{\text{T}}} - (Q \circ B)^{\text{T}}. \tag{2.15}$$

Though its structure is similar to the graph Laplacian in equation (2.9), here it is now a first-order advective operator with directionality. The diagonal matrix $D_{Q^{\mathrm{T}}}$ records the total incoming (upstream) flux into $v_i$. The off-diagonal terms $Q \circ B$ (element-wise matrix product) store the incoming fluid flux $Q$ at each upstream pore outlet, multiplied by the factor $B$ defined as

$$B_{ij}(T) := \exp\left(-\frac{\pi\Lambda \int_0^{L_{ij}} R_{ij}(Z,T)\, dZ}{Q_{ij}}\right). \tag{2.16}$$

This factor represents the proportion of concentration lost to the pore wall due to foulant deposition (see equation (2.12)). Solving the linear system in equation (2.13) yields the particle concentration $C_i$ at each vertex $v_i \in V \backslash V_{\mathrm{in}}$.

The pore (edge radius) kinematics follows a model (see, e.g. Sanaei & Cummings (2017), among others) such that the rate of change of edge radius is directly proportional to local particle concentration,

$$\frac{\partial R_{ij}}{\partial T} = -\Lambda\alpha C_{ij}(Y,T), \quad R_{ij}(Y,0) = R_n, \quad (v_i, v_j) \in E_n, \quad n = 1,\ldots,m, \tag{2.17}$$

where $\alpha$ is a constant related to foulant particle volume. In our model, we also assume that all radii within the $n$-th layer initially take the same value.

Equation (2.17) closes the membrane filtration model with adsorption and, when solved numerically, provides an accurate solution that obeys mass conservation at all points along the pore wall. However, solving the full model for the numerically-determined $R_{ij}(Y,T)$ for every pore in a large network is computationally prohibitive, requiring repeated quadratures. We instead assume that the pore radius satisfies a linear relationship in space, i.e.,

$$R_{ij}(Y,T) = a(T) + b(T)\,Y.$$

Finding the radius of any pore in the network now reduces to solving ODEs for the coefficients $a(T)$ and $b(T)$, derived in appendix A. This approximation yields a slight underestimate of the conductance of the pore, and only a small quantitative change in behavior of the entire pore network, while significantly boosting computational efficiency; a key advantage of this simplification is that spatial quadrature in equation (2.12) can be avoided since the integral of a linear function can be expressed analytically in terms of the coefficients.

### 2.4. *Sieving*

In this section, we highlight the focus of the paper: the incorporation of sieving as a simultaneous fouling mechanism. Sieving is a fouling process explored in many filtration scenarios involving particles as large as a typical pore size (simply, *sieving particles*, hereafter), and the time scale for individual sieving events is much shorter than that of adsorption. Sieving manifests in two primary types, which describe the extent of blockage. The first type is *complete blocking*, when sieving particles completely cover an inlet or clog a pore throat in the interior of the membrane (see figure 1 for the asterisks mimicking sieving particles), prohibiting local fluid flow. The second type is *incomplete blocking*, when sieving particles partially cover the entrance of a pore throat and thus impede fluid flow by significantly narrowing the entrance size (feed solution leaks through the partially blocked channel but at a considerably lower volumetric flow rate). For simplicity, we consider complete blocking only in this work.

### 2.4.1. *Assumptions*

We start with some general assumptions about the sieving process and then provide the proper definitions to set up the numerical simulation. We will continue to use the terminology defined for the graph $G$ in section 2, wherever applicable.

Sieving particles in the feed solution are assumed to arrive at the membrane top surface following a Poisson process $N(T)$ with rate $\Gamma$, i.e.

$$\text{Prob}\,(N(T) = k) = \frac{(\Gamma T)^k}{k!} e^{-\Gamma T}, \quad k = 0, 1, 2 \dots . \tag{2.18}$$

where Prob $(N(T) = k)$ is the probability of having $k$ arrivals over a time span of $T$. We consider monodisperse sieving particle radius $P_{\text{size}}$ in the feed solution, with a primary goal of studying how the particle size affects performance of filters with different pore-radius gradients.

When the $k$-th particle of radius $P_{\text{size}}$ arrives at the entry of a particular edge $e_{ij}$ at time $T^{(k)}$, the principle of *size exclusion* is invoked: if $P_{\text{size}} < R_{ij}$, the particle will immediately pass through and arrive at the vertex $j$ (pore junction). The subsequent direction of particle movement at the junction is governed by *preferential flow* (defined more precisely in section 2.4.2) — the larger the flux in an edge, the more likely the particle will go there. A particle will continue to travel in this fashion through the membrane until it either enters an edge with a radius smaller than or equal to its size (size exclusion) and thereby blocks the pore, or exits the membrane. Lastly, we also assume that sieving is irreversible, i.e., once an edge is blocked, it cannot be unblocked.

### 2.4.2. *Probabilistic Setup*

We simulate blocking by first generating Poisson arrival times based on the fact that inter-arrival times (of particles), $T^{(k+1)} - T^{(k)}$, are exponential random variables with mean $1/\Gamma$. At each arrival time $T_k$, a particle arrives at the top surface and chooses an inlet (any $v_i \in V_{\text{in}}$) with probability proportional to the flux through that inlet relative to the total flux from all $v_i \in V_{\text{in}}$. The size exclusion principle is then applied to the sieving particle to determine which edges it can enter next. If the particle can enter an edge, it will travel to one of the neighboring downstream pores via the edge that connects them, with probability determined by *preferential flow*, defined below. If the particle cannot enter any downstream edges then it will block an edge, again with probability determined by preferential flow.

Definition 2.2. *(Preferential flow) A sieving particle arriving at the membrane top surface has a probability of visiting vertex $v_i$*

$$\pi_{0,i} = \begin{cases} \dfrac{\sum_{v_j \in \mathcal{N}(i)} Q_{ij}(T)}{\sum_{v_i \in V_{\text{in}}} \sum_{v_j \in \mathcal{N}(i)} Q_{ij}(T)}, & \forall v_i \in V_{\text{in}}; \\ 0, & \text{otherwise}, \end{cases} \tag{2.19}$$

*and the probability of moving from $v_i$ to $v_j$ (via $e_{ij}$) is defined by the transition probability*

$$\mathcal{P}_{ij}(T) := \begin{cases} \dfrac{Q_{ij}(T)}{\sum_{v_j \in \mathcal{N}(i)} Q_{ij}(T)}, & \forall v_i \in V \backslash V_{\text{out}} \quad (\textit{flux-weighted random walk}); \\ \delta_{ij}, & \forall v_i \in V_{\text{out}} \quad (\textit{absorbing at bottom surface}), \end{cases} \tag{2.20}$$

*where $Q_{ij}(T)$ is the fluid flux that goes from vertices $v_i$ and $v_j$ through the edge $e_{ij}$, $\mathcal{N}(i)$ the set of neighbors adjacent to $v_i$, and $\delta_{ij}$ the Kronecker delta.*

The initial distribution $\pi_0$ of the random walker (representing the sieving particle) is the probability of entering the membrane at each vertex on the membrane surface (and zero probability of visiting any other vertices). It is quantified by the proportion of outgoing flux at each vertex on the membrane surface relative to the total outgoing flux from all vertices on the membrane surface. The matrix $\mathcal{P}$ is referred to as the normalized in-degree adjacency, representing the transition law for each sieving particle following the preferential flow (weighted by fluid flux). Note that the neighbor set $\mathcal{N}(i)$ over which we sum the $i$-th row of $Q$ works consistently with the definition of flux-weighted adjacency matrix $Q$ as it contains only positive entries for flux to go from an upstream pore to a downstream one (while the reverse direction incurs zero flux). Sieving particles visiting and making it through (per the *size exclusion* mechanism described in section 2.4.1) the outgoing vertex set $V_{\text{out}}$ will be absorbed there, a condition given by the second line of equation (2.20).

Under both adsorption and blocking, the conductance of the edge $e_{ij}$ is given by

$$K_{ij}(T) = \frac{\pi R_{ij}^4(T)}{8\mu L_{ij}} \mathbf{1}_{\{e_{ij}\ \text{open}\}}(T), \quad 0 \le T \le T_{\text{final}}, \tag{2.21}$$

where the indicator function serves as a switch between an open and a blocked pore (due to the instantaneous nature of sieving).

In simulating filtration through a network, we impose a *stopping criterion* that membrane filtration ends either when total flux is less than some prescribed small tolerance, or when there exist no flow paths between any vertices in $V_{\text{in}}$ and $V_{\text{out}}$, due to pore closures. A path-finding algorithm checks the latter criterion. We terminate the filtration at the earliest time $T_{\text{final}}$ when either criterion is satisfied.

## 3. Measures of Performance

Volumetric throughput of a membrane filter over its lifetime is a standard measure of performance in filtration applications (see Pieracci *et al.* (2018), as an example in industrial practice, among all other references in this paper involving filtration and its mathematical models).

Definition 3.1. *The volumetric throughput $H(T)$ through the filter is defined by*

$$H(T) = \int_0^T Q_{\text{out}}(T')\,dT', \quad Q_{\text{out}}(T) = \sum_{v_j \in V_{\text{bot}}} \sum_{v_i:(v_i,v_j)\in E} Q_{ij}(T),$$

*where $Q_{\text{out}}(T)$ is the total volumetric flux leaving the filter.*

In particular, we are interested in $H_{\text{final}} := H(T_{\text{final}})$, the total volume of filtrate processed by the filter over its lifetime.

Another performance metric is the accumulated foulant concentration in the collected filtrate at the membrane outlet, which captures the efficiency of aggregate adsorptive filtration.

Definition 3.2. *The accumulated foulant concentration is defined by*

$$C_{\text{acm}}(T) = \frac{\int_0^T C_{\text{out}}(T')\,Q_{\text{out}}(T')\,dT'}{\int_0^T Q_{\text{out}}(T')\,dT'},$$

*where*

$$C_{\text{out}}(T) = \frac{\sum_{v_j \in V_{\text{bot}}} \sum_{v_i:(v_i,v_j)\in E} \boldsymbol{C}_j(T)\,\mathbf{Q}_{ij}(T)}{Q_{\text{out}}(T)}.$$

## 4. Nondimensionalization

We nondimensionalize the model with the following scales,

$$
\begin{gathered}
\left(D_{\max}, D_{\min}, L_{ij}, R_{ij}, P_{\text{size}}, Y\right) = W\left(d_{\max}, d_{\min}, l_{ij}, r_{ij}, p_{\text{size}}, y\right), \\
\left(T, T^{(k)}, T_{\text{final}}\right) = \frac{W}{\Lambda \alpha C_0}\left(t, t^{(k)}, t_{\text{final}}\right), \\
P = P_0 p, \qquad Q_{ij} = \frac{\pi W^3 P_0}{8\mu} q_{ij}, \qquad K_{ij} = \frac{\pi W^3}{8\mu} k_{ij}, \qquad k_{ij} = \frac{r_{ij}^4}{l_{ij}}, \\
\left(C, C_{\text{out}}, C_{\text{acm}}\right) = C_0\left(c, c_{\text{out}}, c_{\text{acm}}\right), \qquad \left(H, H_{\text{final}}\right) = \frac{W^3}{\alpha C_0}\left(h, h_{\text{final}}\right), \\
\Lambda = \frac{W P_0}{8\mu}\lambda, \quad \Gamma = \frac{|E|\,\Lambda \alpha C_0}{W}\gamma,
\end{gathered}
\tag{4.1}
$$

where $|E|$ is a typical number of pores in a network, usually at the order of $10^4$ via our network generation protocol. One can view $\gamma$ as the dimensionless arrival rate of large particles **per pore** (relative to the rate at which adsorptive fouling occurs).

Under these scalings, we obtain a dimensionless linear system for pressure $p$ and flux $q$,

$$L_k p = 0, \tag{4.2a}$$

$$p(v) = 1, \quad \forall v \in V_{\text{in}}; \quad p(v) = 0, \quad \forall v \in V_{\text{out}}, \tag{4.2b}$$

$$q_{ij} = k_{ij}\left(p(v_i) - p(v_j)\right), \quad \forall \left(v_i, v_j\right) \in E, \tag{4.2c}$$

where $L_k$ is defined in equation (2.9); for Poisson arrivals, we have

$$\text{Prob}\left(N(T) = k\right) = \frac{(\gamma\,|E|\,t)^k}{k!} e^{-\gamma|E|t}, \quad k = 0, 1, 2 \ldots; \tag{4.3}$$

for foulant concentration $\boldsymbol{c}$,

$$L_Q^{\text{in}} \boldsymbol{c} = (q \circ b)^{\mathrm{T}} \boldsymbol{c}_0, \quad L_Q^{\text{in}} = D_{q^{\mathrm{T}}} - (q \circ b)^{\mathrm{T}}, \tag{4.4a}$$

$$\boldsymbol{c}_0 = (1, \ldots, 1, 0, \ldots, 0)^{\mathrm{T}}, \quad b_{ij} = \exp\left(\frac{-\lambda \int_0^{l_{ij}} r_{ij}(y, t)\, dy}{q_{ij}}\right), \tag{4.4b}$$

where $\mathbf{L}_Q^{\text{in}}$ is given by equation (2.15); and for pore radius $r_{ij}$ (for the pore $e_{ij} = (v_i, v_j)$),

$$\frac{\partial r_{ij}(y, t)}{\partial t} = -\boldsymbol{c}_{ij}(y, t), \quad r_{ij}(0) = r_0, \quad \forall \left(v_i, v_j\right) \in E. \tag{4.5}$$

The dimensionless throughput is given by

$$h(t) = \frac{1}{\lambda} \int_0^{\mathrm{t}} q_{\text{out}}(t')\, dt', \quad q_{\text{out}}(t) = \sum_{v_j \in V_{\text{out}}} \sum_{v_i : \left(v_i, v_j\right) \in E} q_{ij}(t), \tag{4.6}$$

and dimensionless accumulated foulant concentration is written as

$$c_{\text{acm}}(t) = \frac{\int_0^t c_{\text{out}}(t')\, q_{\text{out}}(t')\, dt'}{\int_0^t q_{\text{out}}(t')\, dt'}, \tag{4.7}$$

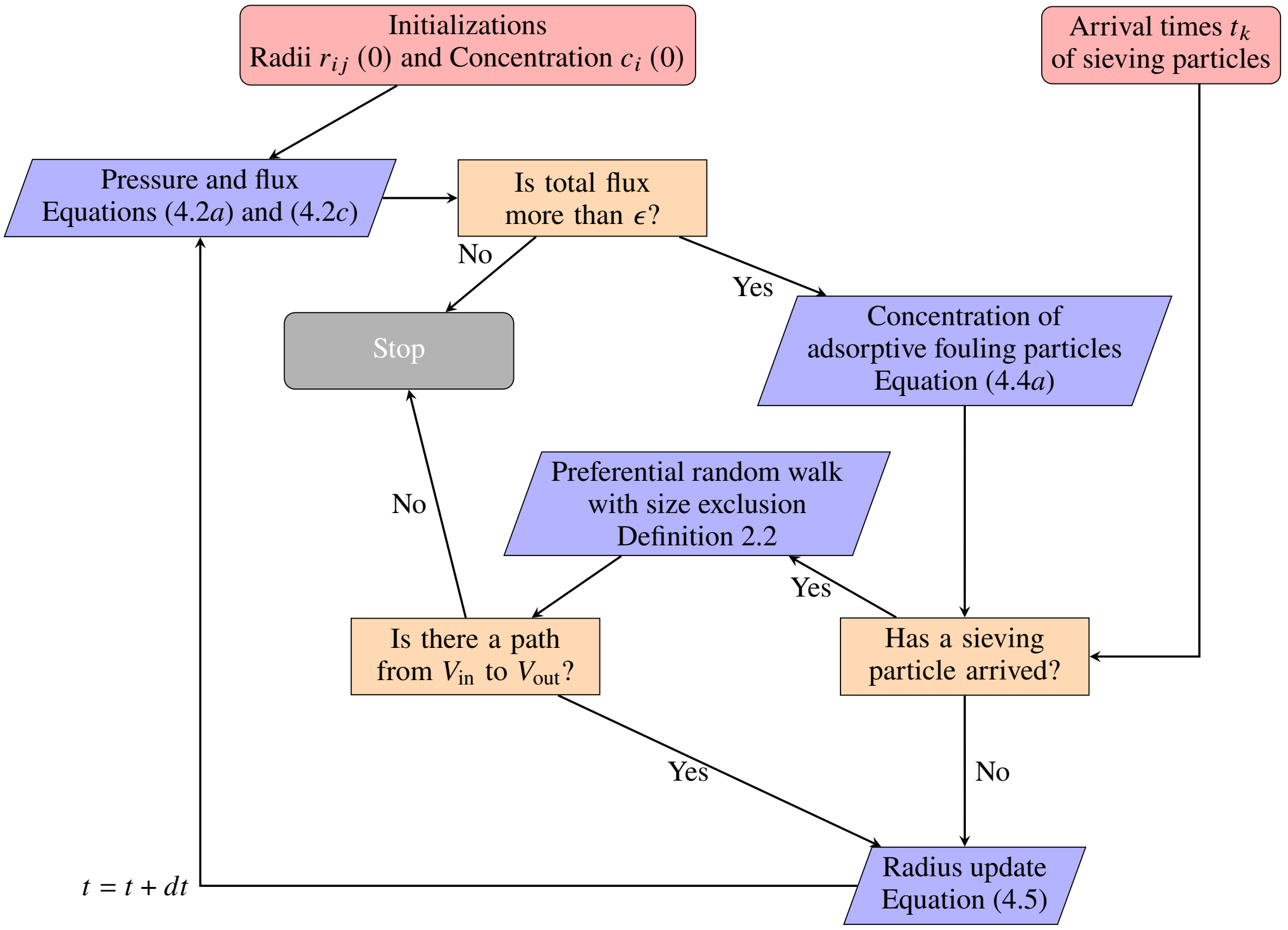

Figure 2. A schematic of the workflow of numerical simulations with simultaneous adsorption and sieving, according to the dimensionless governing equations in section 4. Boxes in pale red represent initialization steps; parallelograms in purple represent numerical equations solving steps; and rectangles in orange represent binary decisions that either trigger a separate calculation or termination of the algorithm.

| | | | |
|---|---|---|---|
| $v_i$ | $i$-th vertex | $e_{ij}$ | An edge connecting vertices $v_i$ and $v_j$ |
| $V$ | Set of all vertices | $E$ | Set of all edges |
| $V_{\text{in}}$ | Set of vertices on the top membrane surface ($\lvert V_{\text{in}}\rvert \approx 750-950$) | $V_{\text{out}}$ | Set of vertices at the bottom membrane surface ($\lvert V_{\text{out}}\rvert \approx 950-2000$) |
| $V_{\text{int}}$ | Set of vertices in the network interior ($\lvert V_{\text{int}}\rvert \approx 1600-1850$) | $m$ | Total number of bands (fixed at 4) |
| $L_{\mathcal{W}}$ | (Weighted) Graph Laplacian of $G$ | $\mathcal{W}$ | Weighted adjacency matrix of $G$ |
| $p_i$ | Pressure at the vertex $v_i$ | $c_i$ | Concentration in vertex $v_i$ |
| $c_{\text{out}}$ | Instantaneous concentration of adsorptive foulant at membrane outlet | $c_{\text{acm}}$ | Accumulated concentration of adsorptive foulant at membrane outlet |
| $k_{ij}$ | Conductance of edge $e_{ij}$ | $q_{ij}$ | Flux through edge $e_{ij}$ |
| $r_{ij}$ | Radius of edge $e_{ij}$ | $l_{ij}$ | Length of edge $e_{ij}$ |
| $d_{\max}$ | Maximal edge length (fixed at 0.15) | $d_{\min}$ | Minimal edge length (fixed at 0.06) |
| $\lambda$ | Adsorption coefficient (fixed at $5 \times 10^{-7}$) | $\gamma$ | Sieving particle arrival rate |
| $\pi_0$ | Initial probability distribution of arrivals on membrane top surface | $\mathcal{P}$ | Transition matrix for preferential flow |

Table 1. Key nomenclature, symbols and values for dimensionless quantities used throughout this work. $\lvert V_{\text{in}}\rvert$, $\lvert V_{\text{out}}\rvert$ and $\lvert V_{\text{int}}\rvert$ depend on the radius gradient $s$ per porosity-fixed network generation protocol described in section 2.1.

where

$$c_{\text{out}}(t) = \frac{\sum_{v_j \in V_{\text{out}}} \sum_{v_i : (v_i, v_j) \in E} c_j(t)\, q_{ij}(t)}{q_{\text{out}}(t)}.$$

See figure 2 for a diagram of the algorithm that simulates these governing equations, and table 1 for a list of key symbols and dimensionless quantities.

## 5. Results

This section presents computational results on the physical output and performance metrics defined in section 4. We first outline the computational setup and parameters employed in the simulations. We emphasize that the goal of this work is to understand the interplay between the new fouling mode of sieving, the pore radius gradients $s$, and the monodisperse sieving particle size $p_{\text{size}}$ at different arrival rates $\gamma$, forming a parameter set $(s, p_{\text{size}}, \gamma)$. Geometric parameters, such as the maximum and minimum pore length ($d_{\max} = 0.15$ and $d_{\min} = 0.06$), the number of bands $m = 4$, and the adsorptive coefficient $\lambda = 5 \times 10^{-7}$, are held fixed. Furthermore, we do not explore the geometric variations of networks due to the random graph generation protocol (discussed in section 2). Instead, we focus on a single realization of network generation for each radius gradient. We have verified (results not shown here for brevity) that outputs from a second, independent realization are very similar. For investigations on geometric variations in setups similar to our work (but without sieving), see Gu *et al.* (2022, 2023).

To explore the effect of pore radius gradient, we consider three cases: an ungraded "uniform" network with $s = 0$, a moderately-graded network with $s = 0.0015$ (as our benchmark network) and a severely-graded network with $s = 0.003$. Since the band porosity is fixed for either network, a gradient in node density is inevitable – the larger the band radius in one network relative to the other, the fewer nodes the band contains. In table 2 we tabulate the radius in each band at each radius gradient. We chose four particle sizes based on their ability to initially sieve at each band in the moderately-graded case. For example, a particle with $p_{\text{size}} = 0.0078$ (the smallest considered in this work) will initially be able to pass through the 3rd band of the moderately-graded network but not the 4th.

We now address the stochastic simulation using equation (4.3). Given the random nature of Poisson arrivals, for each parameter triple $(s, p_{\text{size}}, \gamma)$, we simulate 120 realizations of the Poisson process and collect statistics (average and standard deviation) on scalar performance metrics such as total throughput $v_{\text{final}}$, accumulated foulant concentration $c_{\text{acm}}$ and filter lifetime $t_{\text{final}}$, and on time series data such as total flux $q_{\text{out}}$ through the filter, pressure, and foulant concentration at each vertex. For each simulation, we generate a sequence of arrival times, with inter-arrival times exponentially distributed with mean $1/\gamma$. In other words, the arrival events are "predetermined" before each simulation begins.

To solve the governing equations numerically, the ODEs in equation (4.5) are solved using the forward Euler method with an adaptive time-stepping scheme. This adaptivity ensures that the timestep is small enough to resolve short inter-arrival times, which is particularly important (and likely) for high arrival rates. When no arrival event occurs, the timestep is fixed at $dt = 10^{-5}$. However, when an arrival event is detected, the timestep is adjusted to $0.9\left(t^{(k+1)} - t^{(k)}\right)$ which represents 90% of the $k$th inter-arrival interval, so as to prevent skipping events during time integration.

Given the inherent randomness of the sieving process, filters with identical input parameters may have different lifetimes. Additionally, the adaptive time-stepping scheme results in time series data from different realizations being recorded at non-uniform time points. Therefore, to ensure meaningful statistical analysis, a careful statistical averaging procedure is necessary. For each parameter set $(s, p_{\text{size}}, \gamma)$, we first determine the maximum filter lifetime, $t_{\max}$, among the 120 samples. Next, for each time series (e.g., total flux $q_{\text{out}}(t)$) corresponding to the given parameters, we define a uniform time grid over $[0, t_{\max}]$ with a mesh size of $10^{-8}$, ensuring sufficient resolution to distinguish inter-arrival times at the highest arrival rate considered in this work with overwhelming probability. The time series data are then linearly interpolated onto this grid. To handle simulations that terminate before $t_{\max}$, the flux and throughput remain constant beyond their termination point, effectively padding the interpolated time series with a fixed value. This approach ensures that all time series associated with a given parameter set are aligned on a common time grid, enabling consistent pointwise averaging over time.

| Parameters symbol | Physical meaning | Values |
|---|---|---|
| $p_{\text{size}}$ | Sieving particle size | $0.0078, 0.0093, 0.0108, 0.0123$ |
| $\gamma$ | Sieving particle arrival rate | $[0, 31]$ |
| $s$ | Radius gradient | $0, 0.0015, 0.003$ |

| Initial Radius | $s = 0$ (Ungraded) | $s = 0.0015$ (Moderate) | $s = 0.003$ (Severe) |
|---|---|---|---|
| Band 1 | 0.01 | 0.01225 | 0.0145 |
| Band 2 | 0.01 | 0.01075 | 0.0115 |
| Band 3 | 0.01 | 0.00925 | 0.0085 |
| Band 4 | 0.01 | 0.00775 | 0.0055 |

Table 2. Upper table: parameter values used in simulations. Lower table: radius in each band corresponding to each radius gradient.

With the computational setup described above, we are now ready to present the results and discuss their implications. All quantities of interest are the sample averages described above, unless otherwise noted.

### 5.1. *General Dynamics*

Before examining performance metrics in detail, we first clarify how blocking affects each of the three pore-radius graded networks. In the ungraded network, blocking always occurs at the top surface, regardless of $p_{\text{size}}$, because adsorptive particles have higher concentration upstream, causing inlet pores to shrink fastest. In contrast, graded networks can exhibit internal blocking: particles small enough to pass through larger top-surface pores may be trapped deeper in the membrane where pores are narrower. Surface blocking in these cases only arises once upstream pores have narrowed sufficiently due to adsorptive fouling. This mechanism leads to more complex dynamics and performance outcomes.

### 5.2. *Flux vs Throughput*

One key trend monitored in membrane filtration applications is the flux versus throughput plot. In figure 3, we demonstrate this relationship for all three pore radius gradients considered, using $p_{\text{size}} = 0.0078$, the smallest sieving particle size. With this $p_{\text{size}}$ value, in both moderately and severely graded networks, the 4th band will begin sieving first, since this particle size can initially pass through the 3rd band in both networks. As upstream pores narrow due to adsorption, each band becomes susceptible to sieving as filtration continues (until a stopping criterion has been reached). However, for the ungraded network, the sequence of events is qualitatively different. Sieving particles escape the membrane initially, and because of a higher concentration of adsorptive foulants at the membrane top surface, the pores here shrink fastest. Blocking, therefore, occurs only at the inlets in the 1st band, at the time when they become smaller than the particle size.

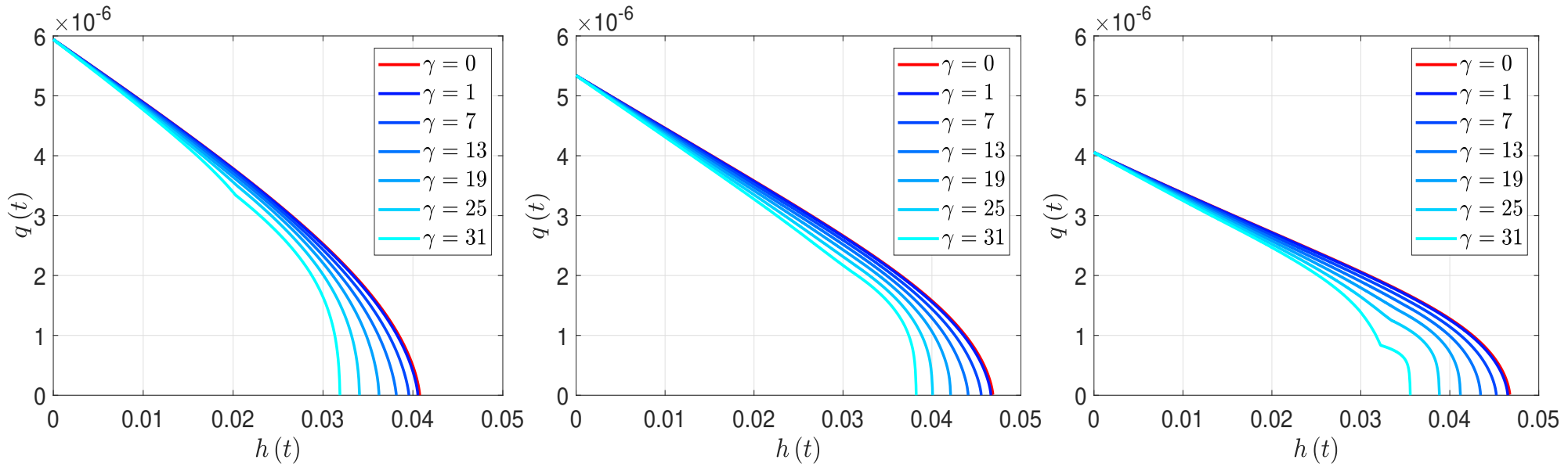

(a) Ungraded (b) Moderate (c) Severe

Figure 3. Flux vs throughtput for networks with pore radius gradient (a) $s = 0$ (ungraded) (b) $s = 0.0015$ (moderate) and (c) $s = 0.003$ (severe). For all plots, $p_{\text{size}} = 0.0078$.

Figure 3 shows results for several different values of the arrival rate $\gamma$, which influences the flux and total throughput obtained. In our benchmark moderately-graded network figure 3b, we observe a consistent decline in both metrics as $\gamma$ increases: flux-throughput curves drop more rapidly, and their $x$-intercepts (total throughput over the filter lifetime) decrease. This is expected because, while adsorption contributes to fouling across all cases, the additional sieving that occurs as $\gamma$ increases accelerates the deterioration of the filter: for higher arrival rates the filter receives a heavier foulant load overall and thus blocks sooner. The same trend is evident in both the ungraded (figure 3a) and severely graded (figure 3c) networks. We also note in passing that the initial flux decreases as a function of radius gradient (see the $y$-intercepts of all subfigures of figure 3). This is because, the larger the gradient, the smaller and more numerous the pores in the bottom band, contributing significantly to the filter's overall (initial) resistance.

For each pore radius gradient, the total throughput does not vary dramatically with the arrival rate (though the ungraded network performs worst at all $\gamma$-values). We attribute the weak influence on total throughput to the small particle size, but for different reasons for each radius gradient case. For the ungraded network, small particles initially pass through the network until the upstream band 1 pores become small enough to block. Flux is unaffected by sieving for most of the filter lifetime, maintaining throughput production reasonably well. However, when sieving begins, the flux decline is more drastic than in other cases because blocking occurs exclusively on the membrane top surface, shutting down the inlets and significantly reducing the number of possible paths from the top surface to the bottom. In essence, the relative gain in throughput due to the initial absence of sieving is mitigated by the catastrophic blocking at the inlets later. This filter pore network performs the worst of the three considered, across all arrival rates. For both of the graded networks, however, blocking occurs initially at the 4th band, obstructing flow and thus filtrate production, but not drastically (pores in band 4 are much smaller and more numerous than those in upstream bands). The upstream pores do not reach the critical size that induces blocking until later than their ungraded counterpart, serving to extend filter lifetime (discussed further later) and promote filtrate production relative to that case.

To further distinguish between the total throughput obtained by the three networks considered, we may study the total throughput values when the sieving-particle arrival rate is large. For example, for $\gamma = 31$, the highest rate considered, we see that the moderately graded network (see figure 3b) produces the most throughput, indicating that for this scenario it is the optimal radius gradient of the three considered (the kink seen in the $\gamma = 31$ curve of figure 3c will be discussed later). This observation is particularly relevant for applications when the sieving particles are small but frequent, for which our results suggest that an intermediate value of pore radius gradient should be used. The same ranking holds for lower arrival rates; however, the difference in total throughput between the two graded cases becomes less pronounced. The ungraded case consistently performs the worst.

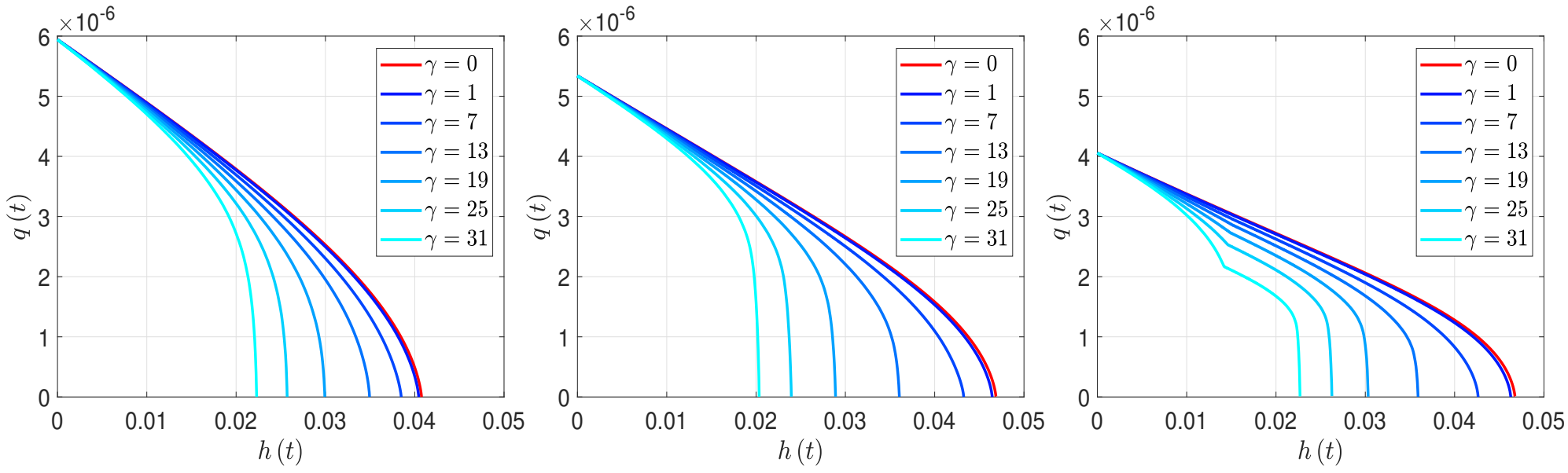

(a) Ungraded (b) Moderate (c) Severe

Figure 4. Flux vs throughput for networks with pore radius gradient (a) $s = 0$ (ungraded) (b) $s = 0.0015$ (moderate) and (c) $s = 0.003$ (severe). For all plots, $p_{\text{size}} = 0.0123$.

In figure 4, we similarly plot flux vs throughput for the same three filters and seven arrival rates as in figure 3, but now using the largest sieving particle size considered, with $p_{\text{size}} = 0.0123$. In each case, sieving begins immediately, but in band 1 for the ungraded and moderately graded networks, and in band 2 for the severely graded network. We still see the expected trend that total throughput decreases as the arrival rate of sieving particles increases, and again, the range of total throughput is only affected by radius gradient to a relatively small degree. However, the dependence on large-particle arrival rate is much stronger here, indicated by the gaps between the flux-throughput curves for different $\gamma$-values being far larger in figure 4 than in figure 3. The reason for this widening gap depends on the radius gradient. For the ungraded and moderately graded cases where sieving starts on the top surface, the number of clogged inlets increases faster as sieving particles arrive more frequently, prohibiting downstream neighboring pores and their subnetworks from producing filtrate (this remains true as long as the sieving particle size is larger than the initial pore size on the top surface). On the other hand, for the severely graded case, the flux-throughput curves separate as a function of arrival rate because the time it takes for the upstream pores to shrink enough to begin sieving is less for the larger particle size. In other words, upstream bands sieve earlier in the filtration process, so that the sieving rate is now a more important factor in reducing flux and total throughput.

For the highest sieving particle arrival rate, $\gamma = 31$, we observe in figure 4 that the moderately-graded filter of figure 4b is the worst in terms of total throughput among the three networks considered, contrary to the results in figure 3b (where it is the best among the three). Combined with the previous observation, this suggests that in applications, filters with moderately graded pore networks should not be used when processing large sieving particles at large arrival rates (in fact, the ungraded and severely graded filters have nearly identical throughputs in this particular case).

Another point to highlight in both figures 3c and 4c is the kink in the flux-throughput curve corresponding to the case of the largest arrival rate $\gamma = 31$ (faintly visible when $\gamma = 25$ as well), a phenomenon absent in the moderately-graded and ungraded cases (see the relatively smoother curves in figures 3a, 3b, 4a and 4b). We discuss this point with reference to the results of figure 3, comparing both moderately and severely graded networks for $p_{\text{size}} = 0.0078$. To relate the time of kink formation to the sieving process, we consider in figure 5 the fraction of blocked pores in each band, defined as

$$\beta_k(t) = \frac{\#\text{ of blocked pores in the } k\text{th band at time } t}{\#\text{ of pores in the } k\text{th band}}. \tag{5.1}$$

This function is monotone non-decreasing since unblocking cannot occur. When it increases, it indicates active sieving in the $k$th band, whereas its flattening signifies cessation of sieving.

In figure 5a, we focus on the fraction of blocked pores in the 1st band as a function of time for the severely graded network and the smallest sieving particle (results of figure 3c). We observe that for the largest arrival rate of sieving particles, $\gamma = 31$, the time of kink formation in the flux curve corresponds to precisely the time at which the first band begins to sieve. Incidentally, this time is also when the second band ceases to sieve, seen in figure 5b by the fraction of blocked pores, $\beta_2(t)$, flattening. One particularly interesting observation lies in figures 5b and 5c, where we see both $\beta_2(t)$ and $\beta_3(t)$ are increasing slightly beyond $t = 0.005$ before both flatten soon after. This shows a momentary but simultaneous cooperation of the 2nd and 3rd bands in their efforts to sieve. This phenomenon is, in fact, singular to the severely graded network when processing small sieving particles because, during the time of cooperation, pore radii in both bands have evolved to be similar in size due to the radius gradient between the two bands. Figure 5d verifies that sieving begins as soon as filtration starts, since the arrival particles are large enough to block the pores in the 4th band of a severely graded network (see table 2 for initial pore radii in each band).

In figure 6, we plot the fraction of blocked pores $\beta_k(t)$ in each band for a moderately graded network using the smallest sieving particle size (results of figure 3b). Here, observing the onset of sieving via the $x$-intercepts of the blue curves through figures 6a to 6d, we find that each band operates sequentially without simultaneous cooperative sieving, i.e., an upstream band must "wait" until its immediate downstream band finishes sieving. The sequential nature of the sieving is because the pore radii in adjacent bands are quite close to each other. Thus, when one band has a pore radius small enough to sieve, its upstream counterparts have also shrunk to a similar size due to adsorption. We verify this claim in figure 6b, where we observe nearly no sieving in the 2nd band at all. In other words, the moderately graded network may not utilize as many pores for sieving as the severely graded one (compare figure 5 and figure 6).

The findings from this section alone show that the severely graded filters perform slightly better than the moderately graded and ungraded counterparts, only in terms of their superior sieving capacity while maintaining a similar amount of filtrate production. The main advantage of severely graded filters, especially when sieving small particles, is the overlap of sieving periods between the interior layers, a feature absent in other networks where only sequential sieving was observed.

### 5.3. *Adsorptive and Sieving Particle Retention*

Though we have shown that the severely graded network is slightly superior in sieving capability and resilience in maintaining total throughput regardless of sieving particle size, we must also assess its efficiency in removing foulants. In this section, we focus on performance metrics that evaluate the accumulated concentration of adsorptive foulant at membrane outlet (ACM in short) and the filter's efficiency in retaining sieving particles.

First, we plot ACM vs throughput in figure 7 for the three pore networks (different pore radius gradients) considered, using the smallest sieving particle size. In all cases, ACM decreases as throughput increases, showing improved adsorptive foulant retention independent of sieving particle arrival rate. Comparing the three figures, we note an obvious hierarchy in ACM, with the ungraded network incurring the largest ACM (dirtiest filtrate) and the severely graded network yielding the least (cleanest filtrate) for each arrival rate. Moreover, figure 7a (ungraded case) demonstrates that total ACM (the $y$-coordinate where each curve ends) is a monotone increasing function of the sieving particle arrival rate. This trend emerges because the intensifying sieving rapidly decommissions pores, effectively retiring them before they have contributed significantly to adsorption. Since adsorptive fouling efficiency depends on the available surface area of the network, this premature loss of active pores due to sieving reduces the network's adsorptive capacity and effectiveness.

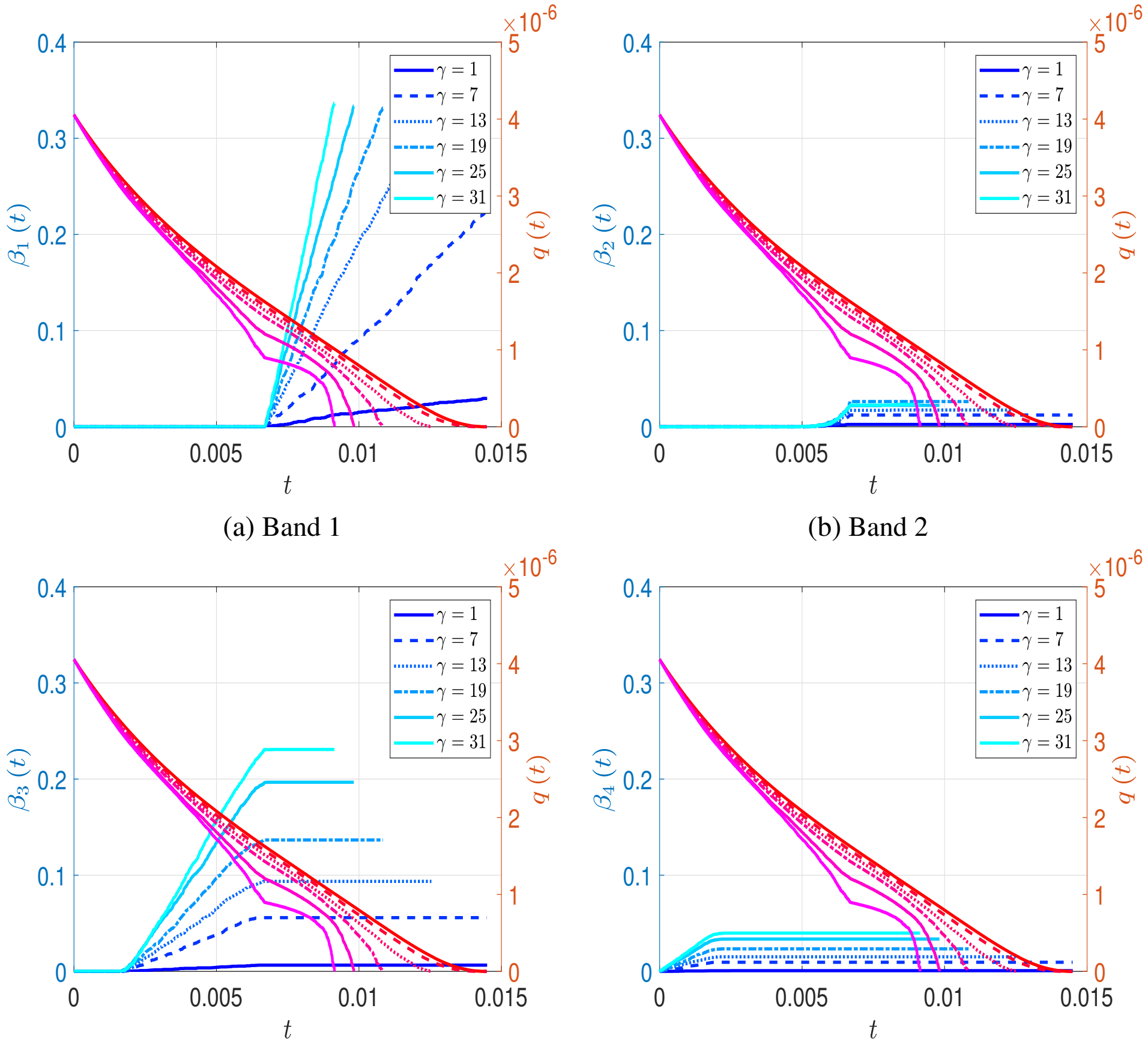

(a) Band 1

(b) Band 2

(c) Band 3

(d) Band 4

Figure 5. Severely-graded networks: blocking percentage and flux against time. The blocked percentage (left $y$-axis, see equation (5.1) for its definition) are in shades of blue, while the accompanying flux (right $y$-axis) are in shades of red. For all plots, $p_{\text{size}} = 0.0078$; results correspond to those of figure 3c.

The dependence of ACM on arrival rate is less pronounced in figure 7b, and barely visible in figure 7c (as the curves collapse), indicating that a nonzero pore size gradient mitigates the loss of adsorption foulant efficiency caused by the competing mode of sieving. This is because the larger upstream pores in these graded networks allow small sieving particles to go deeper into the filter, preserving the upstream pore space for adsorption. Even as sieving arrival rate increases, pore loss is largely confined to smaller downstream (clogged) pores which contribute less to the loss of surface area for adsorption. In essence, the graded architecture induces a division of labor: upstream pores help with adsorption while downstream pores sieve. The radius gradient partly decouples the two fouling modes, reducing the negative interactions observed in the ungraded case.

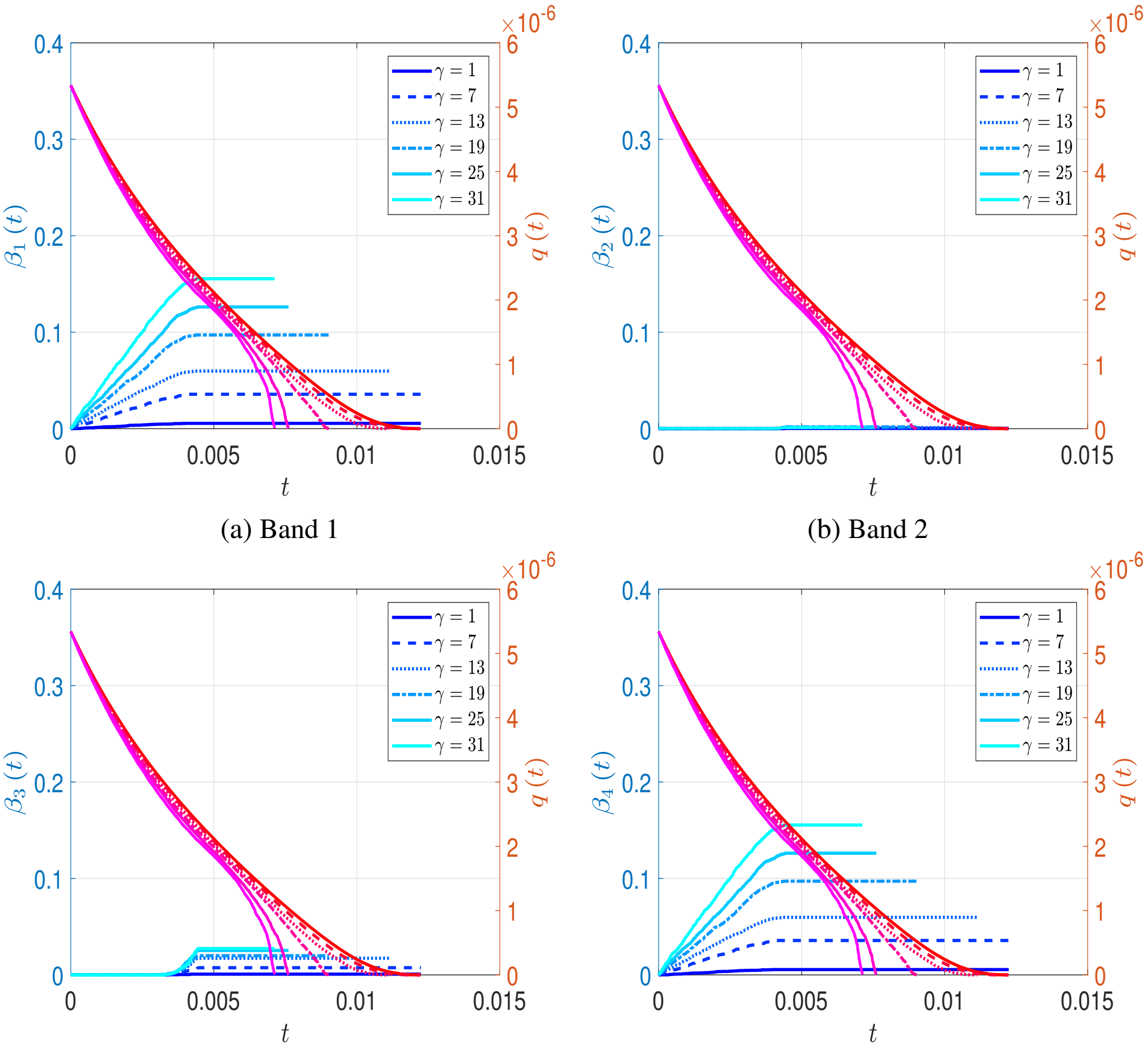

(a) Band 1 (b) Band 2

(c) Band 3 (d) Band 4

Figure 6. Moderately-graded networks (corresponding to results of figure 3b): blocking percentage and flux against time, otherwise the same setup as figure 5.

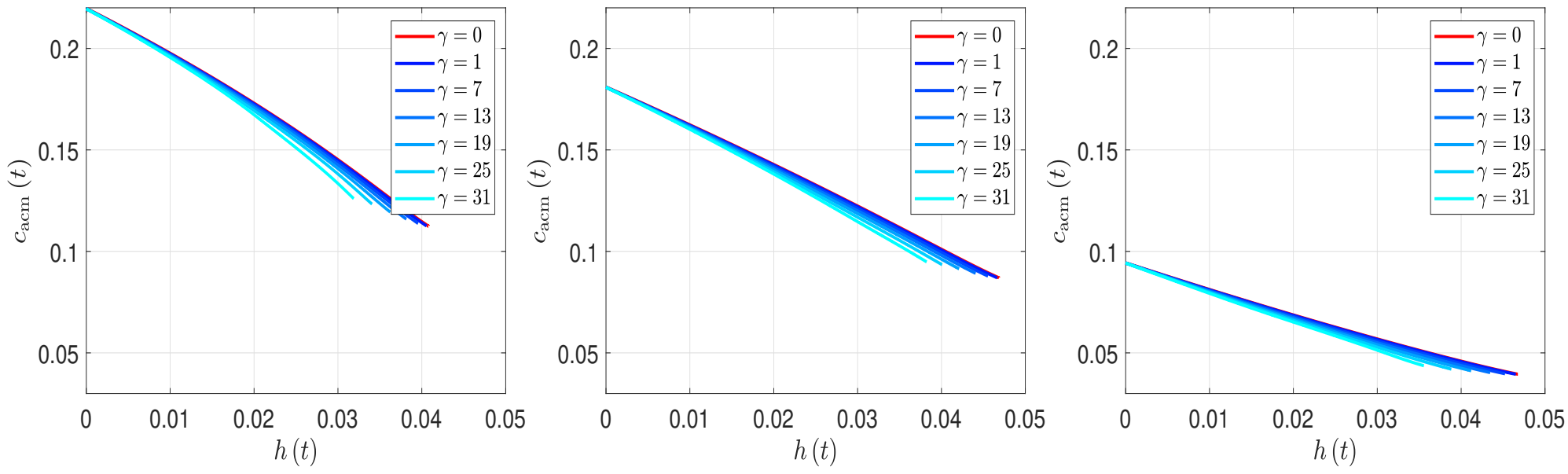

(a) Ungraded (b) Moderate (c) Severe

Figure 7. Accumulated foulant concentration at membrane outlet vs throughtput for networks with pore radius gradient (a) $s = 0$ (ungraded) (b) $s = 0.0015$ (moderate) and (c) $s = 0.003$ (severe). For all plots, $p_{\text{size}} = 0.0078$.

Next, we study ACM vs throughput using the largest sieving particle arriving at several rates in figure 8. The observation from figure 7a, that total ACM is an increasing function of arrival rate, is still evident in figure 8a. Moreover, the gaps between the curves show the intensifying negative impact of increasing arrival rate on total throughput (an observation also discussed earlier in figure 4a). Meanwhile, in figure 8b, we see that the moderately graded network suffers a similar rapid increase in ACM as a function of arrival rate, a feature absent when processing smaller sieving particles (see figure 7b). In figure 8c, for the severely graded network, we see that total ACM is the lowest (the best) among all three considered networks, and here is in fact a *nonmonotone* function of arrival rate, though it varies only a little for all arrival rates considered. More precisely, the severely graded network maintains its adsorption particle retention as the arrival rate increases and even improves as $\gamma = 25 \rightarrow 31$. Curiously, the added mode of sieving appears to benefit the efficiency of adsorptive retention. At the same time, the severely graded network has a relatively similar total throughput to the moderately graded and ungraded cases for each arrival rate.

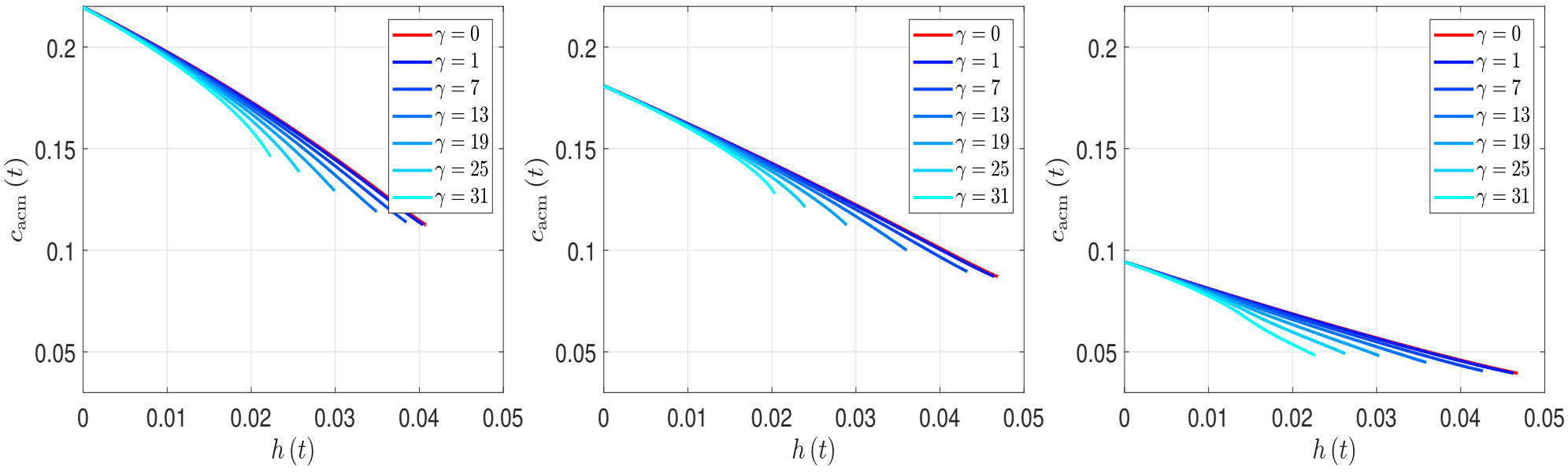

Figure 8. Accumulated foulant concentration at membrane outlet vs throughtput for networks with pore radius gradient (a) $s = 0$ (ungraded) (b) $s = 0.0015$ (moderate) and (c) $s = 0.003$ (severe). For all plots, $p_{\text{size}} = 0.0123$.

Altogether, the results in this section have shown that the severely graded pore network, in terms of adsorptive foulant control and total throughput, is superior to the moderately graded and ungraded networks. The severely graded network is able to withstand frequent arrivals of sieving particles (large $\gamma$) for the smallest and largest $p_{\text{size}}$, respectively, by maintaining a comparable total throughput production to the other two cases, while achieving a much smaller accumulated adsorptive foulant concentration at its membrane outlet. We omit results for intermediate values of $p_{\text{size}}$ for brevity, as the corresponding curves (flux-throughput and ACM-throughput) consistently fall between those obtained by $p_{\text{size}} = 0.0078$ and $p_{\text{size}} = 0.0123$, with intermediate gap sizes (between curves at different arrival rates) reflecting the monotonic trend.

### 5.4. *Filter Lifetime*

In this section, we study how the lifetime of a pore-size graded network depends on model parameters, with reference to the results shown in figure 9. We focus first on the ungraded case, figure 9a, as an illustrative example to develop physical intuition for the system dynamics. Across all values of $p_{\text{size}}$ and network types, filter lifetime generally decreases as $\gamma$ increases, as expected. However, for a given particle size, there appears to be a critical arrival rate above which differences in filter lifetime begin to emerge. To understand this, observe that with no sieving, $\gamma = 0$, adsorption is the only fouling mechanism; and the membrane pore inlets shrink at a unit dimensionless rate, corresponding precisely to a filter lifetime of 0.01 (since pore inlets have an initial radius 0.01, see equation (4.5)). Since pore inlets on the membrane top surface will always close under absorption in the same amount of time, 0.01 represents an upper bound for the filter lifetime (in the ungraded case). For the filter lifetime to decrease from this value, there must be sufficient sieving to close all paths through the pore network before this time; for the ungraded network, the critical arrival rate appears to happen at $\gamma \approx 10$. As $\gamma$ increases beyond 10, filter lifetime exhibits distinct decay rates that depend on the particle size, with larger sieving particles leading to faster decay in filter lifetime. We attribute this distinction to the increasing difference between particle size and the membrane inlets. Filters receiving larger sieving particles begin to foul earlier as pore sizes at the membrane top surface become small enough to cause sieving. This observation explains the emerging difference in filter lifetime between $p_{size} = 0.0078$ and $p_{size} = 0.0093$, the two sizes that are initially passing through the filter uncaptured. On the other hand, the curves for both $p_{size} = 0.0108$ and $p_{size} = 0.0123$ overlie each other since particles of these sizes are capable of sieving at the start of filtration.

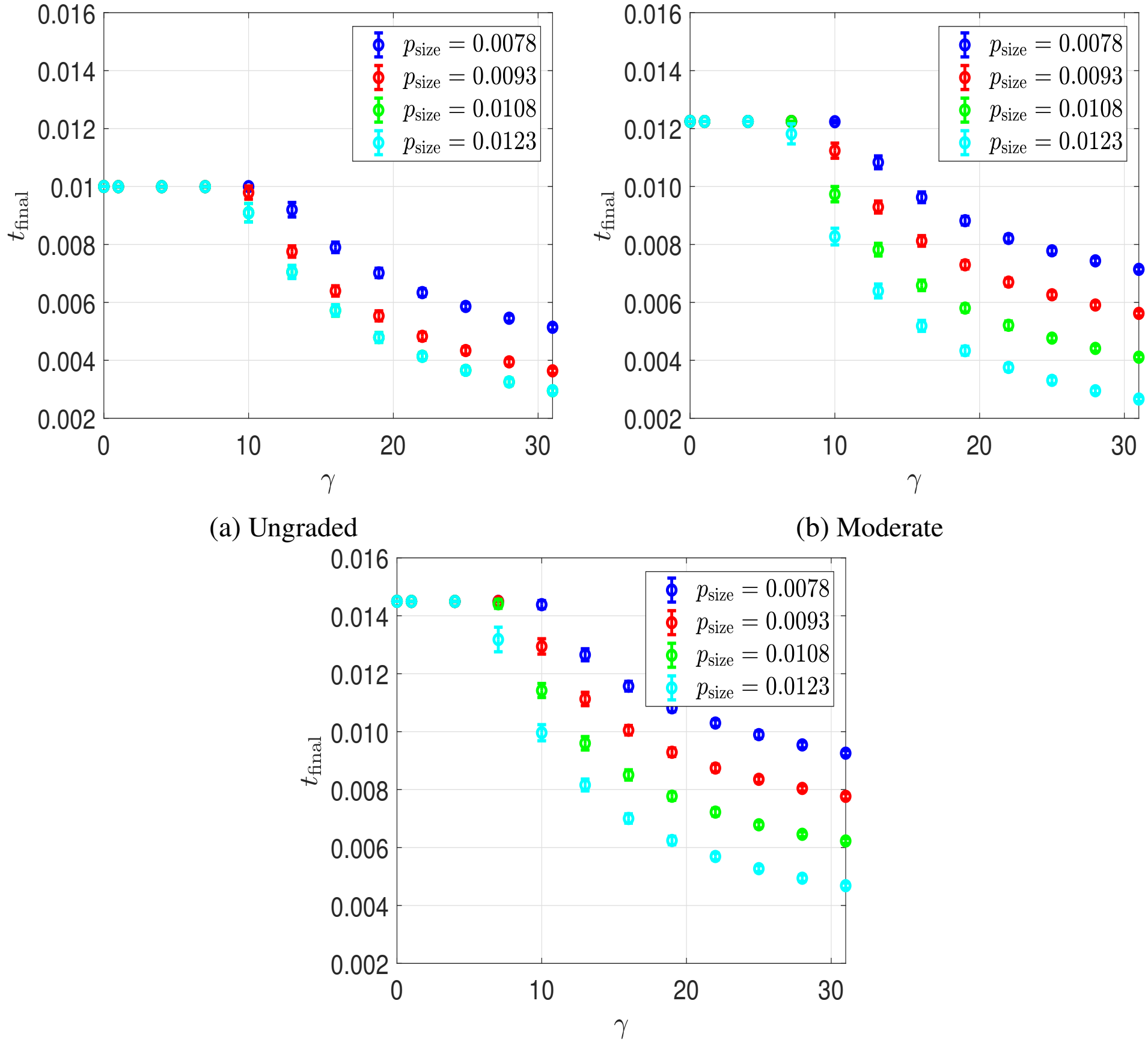

(a) Ungraded (b) Moderate

(c) Severe

Figure 9. Filter lifetime for networks with pore radius gradient (a) $s = 0$ (ungraded) (b) $s = 0.0015$ (moderate) and (c) $s = 0.003$ (severe).

In figures 9b and 9c, we show the corresponding results for the moderately and severely graded pore networks. An important highlight is that given the particle sizes we chose (see table 2), both of these networks are capable of causing sieving in the interior of the network, as opposed to the ungraded case where blocking takes place only at the membrane top surface. A similar phase transition still occurs at some critical arrival rate (though different for each network). When the arrival rate is below the respective critical value for each network, adsorption still dominates, and the filter lifetime is determined precisely by the time when the inlet radii go to zero. The severely graded filter lasts longer in this case, since it has larger upstream pores due to a fixed porosity (compare the $t_{\text{final}}$ values for figure 9b and figure 9c). At the critical arrival rate (the value of which, in contrast to the ungraded case just discussed, now depends quite strongly on $p_{\text{size}}$) the difference in particle sizes begins to play a role. Consider $\gamma = 10$ in figure 9b for example. The four particle sizes contribute to four distinct filter lifetimes due to their respective capabilities of sieving initially at each layer. In all cases, beyond the critical arrival rate the filter lifetime is a monotone decreasing function of particle size. As discussed, this can be attributed to the more efficient distribution of sieving events throughout the network interior for the smaller particles.

## 6. Discussion, Conclusion and Outlook

In conclusion, we have developed a model for membrane filtration in pore-size-graded networks, with a particular focus on the interplay between two simultaneous fouling mechanisms: adsorption due to small particles and sieving of large particles. By introducing a stochastic framework that combines discrete particle-level interactions with network-scale transport dynamics, we are able to describe how sieving particles traverse the membrane and/or block pores, and how network permeability decreases due to both fouling mechanisms over time. Our simulations highlight distinct regimes of filter performance depending on the pore size gradient, sieving particle size, and large particle arrival rate, with the potential for abrupt transitions in fouling patterns and filtration performance metrics. The approach offers a physically interpretable and computationally efficient alternative to direct numerical simulations of full-order nonlinear PDEs, providing insight into critical design trade-offs in graded membrane architectures.

The main findings of this study address three key design questions commonly posed in the development of membrane filters, all framed around a single structural variable—the pore size gradient: regardless of flow conditions, how should pore sizes (radii in our model) be distributed through the depth of the membrane to (1) maximize filtrate volume, (2) minimize foulant accumulation in the filtrate, and (3) prolong filter lifetime? We find that severely graded networks outperform more uniform configurations with respect to the latter two criteria, while offering comparable total throughput. A large pore radius gradient—i.e., a substantial difference between the initial radii in each layer of the filter—ensures that bands further downstream remain available to adsorb foulants and contribute to flux before reaching a size small enough to initiate sieving. This buffering effect is diminished for shallow pore-size gradients and is entirely absent in filters with uniform pore size. Additionally, under the constraint of fixed initial porosity, severely graded networks feature larger upstream pores, which are more resistant to early clogging, thereby extending the operational lifespan of the filter.

While our investigations focused on three representative gradients, we anticipate the existence of an optimal radius gradient that simultaneously maximizes throughput, minimizes foulant accumulation, and extends filter longevity. If the radius gradient becomes excessively large, the pores in the deepest band may become so constricted that, under fixed pressure conditions, negligible flux passes through them—diminishing the filter's effectiveness. This trade-off aligns with several findings in the literature. For example, Griffiths *et al.* (2016) identified a throughput-maximizing tapering angle—an analog to the pore radius gradient; Onur *et al.* (2018) showed that high initial flux often results in poor foulant control due to deposition deep within the filter, as seen in our ungraded networks; and Dalwadi *et al.* (2015) suggested that having more constricted void space downstream improves foulant retention, albeit in the context of porosity gradients rather than pore radius gradients. However, in contrast to these prior studies, our model captures the interaction between sieving and adsorption as coexisting filtration modes, revealing how a judiciously chosen pore radius gradient can help optimize the tradeoff between the two fouling modes.

While the pore-radius-graded networks used in this study yield physically meaningful numerical predictions, future work will focus on improving their relevance to practical applications and alignment with experimental data. Real membrane materials, such as ceramic filters, exhibit significantly greater heterogeneity than is captured by our simple network generation algorithms. For instance, ceramic membranes often possess broad and irregular pore-size distributions that cannot be accurately reproduced by naïve methods. To this end, Shi & Janssen (2025) have employed stratified random sampling to generate volume-weighted pore radius distributions and Sobol sequences to model the spatial arrangement of pore junctions, offering a promising path toward more faithful computational reconstructions. Moreover, while our work highlights how structural features of pore networks can be tuned to manage concurrent sieving and adsorption–and thereby extend filter lifetime–complementary strategies have also explored modifying flow conditions to mitigate clogging. For example, Dincau *et al.* (2022) study pulsatile flow in a microfluidic device with parallel channels, showing that time-dependent driving can delay blockage and prolong filter operation. Incorporating such flow control mechanisms into network-based models like ours presents an intriguing direction for future work, especially in systems where structure and driving dynamics may be jointly optimized.

A key numerical challenge lies in accounting for stochasticity in the sieving particle arrival process, modeled here via Poisson experiments. Although our use of 120 simulations provides a statistically sufficient sample size for this study, a more rigorous statistical analysis is needed to assess convergence and variability in key metrics. In parallel, we have begun exploring a mean-field approximation of the blocking process, wherein each pore evolves according to an average state governed by its time-dependent blocking probability. This probability can be computed using a system of Kolmogorov-type ODEs (under the Poisson measure), which bypasses the need for repeated stochastic realizations and opens the door to investigating alternative sources of randomness, such as uncertainty in pore size distributions, within a deterministic framework. We intend to develop this method further in future work.

Another possible direction involves extending the model to account for capillary effects, particularly relevant in dry porous media prior to the start of filtration. In such settings, the introduction of fluid and foulants triggers phenomena including capillary pressure buildup, air compression, and spontaneous imbibition (Li *et al.* 2017). Capillary network models, such as those studied by Markicevic & Navaz (2010, 2009); Markicevic *et al.* (2011), provide a natural foundation for incorporating these dynamics. Capturing the early-stage saturation process may require solving Richards' equation to describe transient fluid invasion before reaching quasi-steady operating conditions (as explored in the present work).

Finally, the modeling framework developed here belongs to the broader class of PDE dynamics on evolving networks, a rapidly growing field of study. Recent theoretical advances – including those by Berkolaiko & Kuchment (2016), Fijavž & Puchalska (2020), Chapman & Wilmott (2021), Böttcher & Porter (2025), and Nachbin (2025), have highlighted the mathematical richness of these problems. On the computational side, numerical platforms such as the QGLAB (Goodman *et al.* 2025) are beginning to make such models tractable at scale. The blend of analytical and data-driven approaches in this field offers exciting possibilities for further generalizing our work to broader classes of (stochastically) dynamic and adaptive filtration systems.

**Declaration of Interests**

The authors report no conflict of interest.

## Appendix A. Linear Approximation of Pore Profile in Space

In this section, we present the details of the process by which we approximate the pore shape as an evolving cross-section of a circular cone. This approximation applies to all edges considered in our networks, but we suppress the $ij$ notation for edges for simplicity.

Consider a linear approximation of the pore radius in the pore axial coordinate $y$,

$$r\left(y,t\right)=a\left(t\right)+b\left(t\right)y, \tag{A 1}$$

where $a\left(t\right)$ and $b\left(t\right)$ are to be determined as follows. The concentration $c\left(y,t\right)$ satisfies an analytical solution (per equation (4.4*a*))

$$c\left(y,t\right)=c_0\left(t\right)\exp\left(-\frac{\lambda\int_0^y r\left(x,t\right)dx}{q\left(t\right)}\right)$$

where the integral now becomes

$$\int_0^y r\left(x,t\right)dx=a\left(t\right)y+b\left(t\right)\frac{y^2}{2}.$$

Knowing that the pore radius evolves according to $\frac{\partial r}{\partial t}=-c\left(y,t\right)$, we have

$$\frac{\partial r}{\partial t}=\frac{da}{dt}+\frac{db}{dt}y=-c_0\left(t\right)\exp\left(-A\left(t\right)y-B\left(t\right)y^2\right)$$

where

$$A\left(t\right)=\frac{\lambda a\left(t\right)}{q\left(t\right)},\quad B\left(t\right)=\frac{\lambda b\left(t\right)}{2q\left(t\right)}. \tag{A 2}$$

Now, we Taylor-expand the exponential up to second order in $y$:

$$\begin{aligned}\frac{da}{dt}+\frac{db}{dt}y&=-c_0\left(t\right)\left[1-\left(A\left(t\right)y+B\left(t\right)y^2\right)+\frac{\left(A\left(t\right)y+B\left(t\right)y^2\right)^2}{2}+O\left(y^5\right)\right]\\&=-c_0\left(t\right)\left[1-A\left(t\right)y+\left(\frac{A^2\left(t\right)}{2}-B\left(t\right)\right)y^2+O\left(y^3\right)\right]\end{aligned}$$

where we find a compatibility condition for the time-dependent coefficients

$$A^2\left(t\right)=2B\left(t\right)\implies\frac{\lambda^2a^2\left(t\right)}{q^2\left(t\right)}=\frac{\lambda b\left(t\right)}{q\left(t\right)}\implies b\left(t\right)=\frac{\lambda a^2\left(t\right)}{q\left(t\right)}.$$

Then,

$$\frac{da}{dt}=-c_0\left(t\right),\quad\frac{db}{dt}=A\left(t\right)c_0\left(t\right)=\frac{\lambda a\left(t\right)}{q\left(t\right)}c_0\left(t\right), \tag{A 3}$$

with analytical solution for $a(t)$,

$$a\left(t\right)=a\left(0\right)-\int_0^t c_0\left(s\right)ds=r_0-\int_0^t c_0\left(s\right)ds, \tag{A 4}$$

since

$$r(y,0) = a(0) + b(0)\,y \implies a(0) = r(0,0) = r_0.$$

We also have

$$\frac{db}{dt} = \frac{\lambda}{q(t)} c_0(t) \left( r_0 - \int_0^t c_0(s)\,ds \right),$$

and thus

$$b(t) = b_0 + \lambda \int_0^t \frac{c_0(s)}{q(s)} \left( r_0 - \int_0^s c_0(t')\,dt' \right) ds. \tag{A 5}$$

Now, since all pores are initially cylinders with radius $r_0$, we must have

$$r_0 = r(y,0) = r_0 + b(0)\,y \implies b(0) = b_0 = 0.$$

Conductance satisfies $k(t) = \frac{1}{I(t)}$ where

$$\begin{aligned} I(t) &= \int_0^l (a(t) + b(t)\,y)^{-4}\,dy \\ &= -\frac{1}{3b(t)} (a(t) + b(t)\,y)^{-3} \Big|_{y=0}^{y=l} \\ &= \frac{1}{3b(t)} \left[ \frac{1}{a^3(t)} - \frac{1}{(a(t) + b(t)\,l)^3} \right]. \end{aligned}$$

Thus,

$$k(t) = \frac{3b(t)}{\frac{1}{a^3(t)} - \frac{1}{(a(t)+b(t)l)^3}} = \frac{3a^3(t)\,b(t)}{1 - \left( \frac{a(t)}{a(t)+b(t)l} \right)^3}. \tag{A 6}$$

We use this formula for conductance (in terms of time), to update flux and concentration per equations (4.2$a$) and (4.4$a$), respectively. Both of these quantities then influence pore shape, whose evolution is then completely governed by $a(t)$ and $b(t)$. This procedure avoids storing $r$ numerically as a function of space, while still providing an accurate yet very efficient representation of pore evolution.